\documentclass{svmult} % We use the Springer-Verlag format

\usepackage{graphicx}        % standard LaTeX graphics tool
\usepackage[bottom]{footmisc}% places footnotes at page bottom

\begin{document}

\title*{Popper's test of Quantum Mechanics}
\author{
Albert Bramon\inst{1}
\and
Rafel Escribano\inst{2}
}
\institute{
Grup de F{\'\i}sica Te\`orica, Universitat Aut\`onoma de Barcelona, 
E-08193 Bellaterra (Barcelona), Spain
\texttt{bramon@ifae.es}
\and
Grup de F{\'\i}sica Te\`orica and IFAE, Universitat Aut\`onoma de Barcelona, 
E-08193 Bellaterra (Barcelona), Spain
\texttt{Rafel.Escribano@ifae.es}
}

\maketitle

A test of quantum mechanics proposed by K.~Popper and dealing with two-particle
entangled states emitted from a fixed source has been criticized by several authors.
Some of them claim that the test becomes inconclusive once all the quantum aspects
of the source are considered. Moreover, another criticism states that the predictions
attributed to quantum mechanics in Popper's analysis are untenable. We reconsider
these criticisms and show that, to a large extend, the `falsifiability' potential of the
test remains unaffected.

\section{Introduction}
\label{intro}
% Always give a unique label
% and use \ref{<label>} for cross-references
% and \cite{<label>} for bibliographic references
% use \sectionmark{}
% to alter or adjust the section heading in the running head
Heisenberg's principle is the key feature of quantum mechanics \cite{GP} and plays 
the central role in many relevant discussions concerning 
counterintuitive quantum phenomena. This is particularly true when
applied to composite quantum systems consisting of two-particle entangled 
states. A well known case is 
the Einstein-Podolsky-Rosen paradox which appeared in 1935 short after a
preliminary and lesser known proposal by Karl Popper \cite{Popper34}. The latter
``Popper's test" of quantum mechanics  has been subsequently reformulated and
improved by Popper himself \cite{Popper34}--\cite{Popper87}, and reconsidered by
several authors \cite{BS}--\cite{Qureshi2}. A look at the most recent of these
papers shows that the controversy is still open. In the present note we 
attempt to clarify some points of such a basic issue.
\par
Following Popper \cite{Popper87}, we consider a source S decaying symmetrically into
pairs of photons or equal-mass particles. In their center-of-mass frame, the two
particles of each pair are assumed to be simultaneously emitted from the origin and
to travel in opposite directions. 
Assume, for concreteness, that their trajectories are contained in the plane defined
by the horizontal $x$-axis and the vertical $y$-axis.
Two vertical screens  are symmetrically placed at
equal distances $d$, left ($L$)  and right ($R$) of the source, thus 
intersecting perpendicularly the
horizontal axis at $x=\pm d$. The left and right screens have  $s_L$- and $s_R$-wide
slits centered  around the horizontal axis at $x= -d$ and $x=+d$,
respectively. Some pairs of the emitted particles will then pass through the slits and
will trigger coincidence detectors placed far away\footnote{
These detectors are distributed in slightly different configurations
according to the various authors. If the distance to the slits is large enough, all the 
various distributions become equivalent.}
and completely 
covering the vertical space behind each slit. The vertical $y$-component of the momentum
of the left- and right-moving  particles, $(k_1 )_{y} \equiv k_{1}$ 
and $(k_2 )_{y} \equiv k_{2}$, can thus be measured
with this experimental setup. The other, non-vertical components 
of $\vec k_{1}$ and $\vec k_{2}$  play a secondary 
role and the discussion thus centers on how the observed distributions for $k_1$ and
$k_2$ depend on the slit widths, $s_L$ and $s_R$. Assume, for instance, that one
measures the $k_1$ and $k_2$ distributions for particle pairs with both members 
passing through equal-width slits, $s_L =s_R = 2a$. Since this amounts to  position 
measurements with $\Delta y_{1,2}\simeq a$, from diffraction  theory or Heisenberg's 
principle one expects $\Delta k_{1,2}\simeq 1/2a$ for the dispersion of the vertical momenta
behind each slit. 
According to Popper's proposal, by two suitable modifications of the previous setup one can
experimentally discriminate between his own (propensity) interpretation of quantum
mechanics and Copenhagen's orthodoxy \cite{Popper87}. 
\par
A first possibility ---case (i)--- consists in considering what happens to 
$\Delta k_2$ when one runs the experiment with the same 
$s_L = 2a$ as before but with a much wider $s_R$, $s_R \gg 2a$. Note that one
still performs a position measurement with $\Delta y_{1}\simeq a$ 
on the particle passing through the left-side slit and that,
because of the entanglement, the right-moving particle has then to be 
in a state with $\Delta y_{2}\simeq a$ when it passes through the 
vertical plane at $x=+d$. According to
Popper \cite{Popper87}, quantum mechanics predicts the same $\Delta k_{2}\simeq 1/2a$
as before, whereas his own (propensity) approach predicts the disappearance of such a
dispersion in $\Delta k_2$.  Moreover ---case (ii)--- one can then narrow the left-slit
width $s_L$ while  maintaining the same wide and fixed $s_R \gg 2a$, as before.
According to quantum mechanics this narrowing of $s_L$ implies larger values for
$\Delta k_{2}\simeq 1/2a$, while the opposite is expected from Popper's approach
\cite{Popper87}.
\par
The proposal summarized in the previous two paragraphs ---a crucial experiment,
according to Popper--- has however been criticized by several authors. All 
these criticisms are based on the fact that the source S cannot be 
exactly localized at the origin and perfectly at rest, as required to 
argue that the two entangled final particles separate from the origin 
strictly with opposite momenta. The undecayed source itself or, better, the
global two-particle final system has to obey Heisenberg's principle and,
accordingly, the vertical components of the CM-position, $(y_1 + y_2)/2$, and total
momentum, $k_1 + k_2$, must satisfy $\Delta (k_1 + k_2) \Delta 
\left[ (y_1 + y_2)/2 \right] \ge 1/2$. Once this constraint is imposed, the
analyses  of Refs.~\cite{BS,CL} ---based on simple and intuitive geometrical
arguments--- claim that Popper's proposal is no longer able to discriminate
between the two approaches. Similarly, the discussion in \cite{Qureshi1,Qureshi2}
---based on a simplified wave function for the two-particle system which obeys
Heisenberg's principle--- claims that for case (ii) standard quantum mechanics
predicts no increase of $\Delta k_{2}$ when narrowing the left-side slit. A claim
which contradicts Popper's original analysis \cite{Popper87} and would make his test
inconclusive as well. We now proceed to discuss that these claims ---based on simple
geometrical arguments and a too naive wave function--- are not completely justified 
and that Popper's test maintains most of its valuable `falsifiability' potential.

\section{Entangled two-particle state}
\label{state}
Consider the following wave function describing the behavior of an entangled 
two-particle system 
\begin{equation}
\label{xx1}
\Psi(y_1,y_2;t)=\int\int dk_1 dk_2 \Psi(k_1,k_2;t)
{e^{i k_1 y_1}\over\sqrt{2\pi}}{e^{ik_2y_2}\over\sqrt{2\pi}}\ , 
\end{equation}
where
\begin{equation}
\label{kk}
\begin{array}{rcl}
\Psi(k_1,k_2;t) 
&=&{1\over\sqrt{\pi\sigma_+\sigma_-}} 
e^{-{1\over 4\sigma_+(t)^2}\left(k_1+k_2\right)^2}
e^{-{1\over 4\sigma_-(t)^2}\left(k_1-k_2\right)^2}\\[2ex]
\qquad
&=&{1 \over \sqrt{\pi\sigma_+\sigma_-}} 
e^{-{1\over 4}\left({1\over\sigma_+(t)^2}+{1\over\sigma_-(t)^2}\right)
\left(k_1^2+k_2^2\right)}
e^{-{1\over 4}\left({1\over\sigma_+(t)^2}-{1\over\sigma_-(t)^2}\right)2 k_1 k_2}\ ,
\end{array}
\end{equation}
${1\over\sigma_\pm(t)^2}\equiv{1\over\sigma_\pm^2} + i{t\over m}$ accounts for
the time evolution along the relevant, vertical $y$-axis, and $m$ is the mass of
each particle. Here and in what follows, integrations extend from $-\infty$ to 
$+\infty$ unless otherwise is stated.
\par
Note that for the global system we have chosen a Gaussian wave packet
\cite{GP} with $\Delta (k_1+k_2) = \sigma_+$. This allows for analytical
computations and at the decay time, $t=0$, one has $\Delta (k_1 + k_2) \Delta  
\left[ (y_1 + y_2)/2 \right] = 1/2$, which is the minimum value 
compatible with Heisenberg's principle; in this sense, our state is 
the closest quantum analog to Popper's original proposal
with a fixed and well localized source.
\par
Note also that we have similarly chosen a Gaussian packet with 
$\Delta |2 k_{1,2}| \simeq \Delta (k_1-k_2) = \sigma_-$ to describe the vertical
spread of the final momenta. Admittedly, this somehow reduces the generality of
our treatment, but our Gaussian choice simplifies the analysis and by no
means precludes the discussion of Popper's proposal which was intended to be valid
for a generic wave packet. Physically, the momentum distribution is isotropic in $s$-wave
decays, such as positronium annihilation into two photons; restricting 
to particles moving in the $xy$-plane, the vertical components
of their momenta, $k_{1,2}$, are uniformly distributed in the
range  $-|\vec k_{1,2}| \le k_{1,2} \le +|\vec k_{1,2}|$. 
For vertically polarized spin-1 states decaying into two
spinless particles, as in vector-meson decays into two pseudoscalar 
mesons, one has 
$k_{1,2} = |\vec k_{1,2}| \cos \theta$, where $\theta$ is the angle between 
$\vec k_{1,2}$ and the $x$-axis, and a vertical momentum distribution peaked 
around $k_{1,2} =0$. Initial sources of higher spin can lead to 
distributions with more pronounced peaks around $k_{1,2} =0$. 
We can somehow mimic this various possibilities 
by a judicious choice of $\sigma_-$ in our Gaussian packet. Note finally that one has 
$\sigma_- \gg \sigma_+$ for any realistic value of $|\vec k_{1,2}|$.

\section{Popper's proposal: case ii)}
\label{case2}
In order to discuss the standard quantum mechanical prediction for 
$\Delta k_2$ when the right-side slit is wide open and the width 
$2a$ of the left-side screen is modified, we need to Fourier transform state 
(\ref{kk}) into 
\begin{equation}
\label{xk}
\begin{array}{rcl}
\Psi(y_1;k_2;t)
&=&\int{dk_1\over\sqrt{2\pi}}{e^{ik_1y_1}}\\
&&\times{1\over\sqrt{\pi\sigma_+\sigma_-}} 
e^{-{1\over 4}\left({1\over\sigma_+(t)^2}+{1\over\sigma_-(t)^2}\right)
\left(k_1^2+k_2^2\right)}
e^{-{1\over 4}\left({1\over\sigma_+(t)^2}-{1\over\sigma_-(t)^2}\right)2k_1k_2}\\[2ex]
&=&{\sqrt{2}\over\sqrt{\pi\sigma_+\sigma_-}}
{\sigma_+(t)\sigma_-(t)\over\sqrt{\sigma_+(t)^2 +\sigma_-(t)^2}}\\[2ex]
&&\times e^{-{1\over{\sigma_+(t)^2+\sigma_-(t)^2}}
\left(\sigma_+(t)^2\sigma_-(t)^2 y_1^2+k_2^2-i(\sigma_+(t)^2-\sigma_-(t)^2)y_1 k_2\right)}.
\end{array}
\end{equation}
From these expressions it is easy to compute the  probability for  
observing the vertical position of the left-moving particle within the range 
$-a \le y_1 \le +a$ allowed by the slit 
in coincidence with a given value, $k_2$, for the vertical component of the momentum
of its right-moving partner. The former measurement requires detecting the left particle 
behind the slit using, for instance, a single detector placed on the 
negative $x$-axis far left of the slit. The measurement of the 
vertical component, $k_{2}$, of the right-side momentum is achieved thanks to the 
distant set of right-side detectors. The quantum mechanical prediction for the spread of
the $k_2$ distribution in these coincidence measurements is then unambiguous:
\begin{equation}
\label{anya}
(\Delta k_2)^2|_{a}= 
{\int dk_2 k_2^2 |\int_{-a}^{+a}dy_{1}\Psi(y_1;k_2;t)|^2
\over \int dk_2 |\int_{-a}^{+a}dy_{1}\Psi(y_1;k_2;t)|^2}\ .
\end{equation} 
\par
We can now consider several values of the left-slit width $2a$. If this is
infinitely narrow,  $2a \to 0$ and $y_1=0$, one easily finds  
\begin{equation}
\label{a=0}
(\Delta k_2)^2|_{a\to 0}=\frac{\sigma_+^2 +\sigma_-^2}{4} 
{1+\left({2\sigma_+\sigma_-\over\sigma_+^2+\sigma_-^2}\right)^2 
\sigma_+^2\sigma_-^2{t^2\over m^2}\over 1+\sigma_+^2\sigma_-^2{t^2\over m^2}}\ ,
\end{equation} 
which decreases from ${1\over 4}(\sigma_+^2+\sigma_-^2)$ at $t=0$ to 
${\sigma_+^2\sigma_-^2\over\sigma_+^2+\sigma_-^2}$ when $t\to\infty$. 
Note that these results hold not only for $y_1=0$ but also for any other precise
localization ($2a \to 0$) at a given $y_1$ of the left-moving particle. 
\par
We next increase the width of the left-side slit to a value $2a$  small enough
to allow for an expansion of the $y_1$-Gaussian. Retaining the first three terms of
the expansion, the quantum mechanical prediction for the $k_2$ distribution turns out
to be 
\begin{equation}
\label{asmall}
(\Delta k_2)^2|_{a}=(\Delta k_2)^2|_{a\to 0}
(1-2 a^2\delta)\ , 
\end{equation}
where
\begin{equation}
\delta=\frac{1}{12}\frac{(\sigma_+^2-\sigma_-^2)^2}{\sigma_+^2+\sigma_-^2}
\frac{1}{1+\sigma_+^2\sigma_-^2{t^2\over m^2}}
\frac{1-\left({2\sigma_+\sigma_-\over\sigma_+^2+\sigma_-^2}\right)^2 
\sigma_+^2\sigma_-^2{t^2\over m^2}}
{1+\left({2\sigma_+\sigma_-\over\sigma_+^2+\sigma_-^2}\right)^2 
\sigma_+^2\sigma_-^2{t^2\over m^2}}\ ,
\end{equation}
which is positive for reasonable values of $t$ and thus $(\Delta k_2)^2|_{a}$ 
decreases when increasing the slit-width $s_{L} =2a$. 
\par
We finally consider the other extreme case $a \to \infty$. From Eq.~(\ref{anya}) one
obtains  
\begin{equation}
\label{a=infty}
(\Delta k_2)^2|_{a\to\infty} 
%&=& {\int dk_2 k_2^2 |\int dy_1 \Psi (y_1, k_2;t)|^2
%\over \int dk_2 |\int dy_1 \Psi (y_1, k_2;t)|^2} \nonumber \\
={\sigma_+^2\sigma_-^2\over\sigma_+^2+\sigma_-^2}\ , 
\end{equation}
which is easily seen to be never larger than all the preceding results, Eqs.~(\ref{a=0}) 
and (\ref{asmall}).
\par
The predictions quoted in the last three paragraphs fully confirm
the quantum mechanical analyses by Popper \cite{Popper82}--\cite{Popper87} on the dependence
of $\Delta k_2$ on the left-side slit width $2a$. The larger (narrower) this width is
chosen, the smaller (wider) is the $k_2$-dispersion $\Delta k_2$. It is easy
to see that our treatment allows to confirm a related analysis by Peres \cite{Peres}
where the single left-side slit is substituted by a double slit thus producing
interference effects on the right-side in coincidence measurements.

\section{Popper's proposal: case i)}
\label{case1}
Let us finally move to the other possibility considered by Popper. In this case, 
one has to fix the width of left-side slit to $s_L = 2a$ and compare 
the predictions for $\Delta k^2_2$ when the right-side slit has the same width, $s_R
= s_L= 2a$, with those from another setup with a wide open right-slit, 
$s_R \gg s_L =2a$. This requires to perform a second Fourier transform of the 
state we are dealing with. It then reads,  
\begin{equation}
\label{yy2}
\Psi(y_1,y_2;t)={\sigma_+(t)\sigma_-(t)\over\sqrt{\pi\sigma_+\sigma_-}}  
e^{-{1\over 4}\left[(\sigma_+(t)^2 +\sigma_-(t)^2)(y_1^2 + y_2^2)+ 
(\sigma_+(t)^2-\sigma_-(t)^2)2y_1 y_2\right]}\ . 
\end{equation}
From this expression it is easy to compute the spreading of the vertical position of
the right-moving particle, $\Delta y^2_2 |_{a}$, when it crosses the
vertical plane $x= +d$ at time $t$. Simultaneously, its left-side partner passes through the window   
$-a \le y_1 \le +a$ allowed by the left-slit and will be detected much later. 
For these left-right coincidence detections, one has 
\begin{equation}
\label{anya2}
(\Delta y_2)^2|_{a}={\int dy_2 y_2^2|\int_{-a}^{+a}dy_{1}\Psi(y_1,y_2;t)|^2 \over 
\int dy_2|\int_{-a}^{+a}dy_{1}\Psi(y_1,y_2;t)|^2}\ .
\end{equation} 
\par
For $a\to 0$ one obtains  
\begin{eqnarray}
\label{a=02}
&&(\Delta y_1)^2|_{a\to 0}\to 0\ ,\\
&&(\Delta y_2)^2|_{a\to 0}=\frac{1}{\sigma_+^2+\sigma_-^2}
\frac{\left(1+\sigma_+^4{t^2\over m^2}\right)\left(1+\sigma_-^4{t^2\over m^2}\right)}
{1+\sigma_+^2\sigma_-^2{t^2\over m^2}}\ge 0\ ,
\end{eqnarray}
with $\Delta y^2_2 |_{a\to 0} = \Delta y^2_1 |_{a\to 0} =0$ only if $t\to 0$ and 
$\sigma_{\pm} \to \infty$, i.e., in the case considered by Popper of a perfectly
localized source ($\sigma_{+} \to \infty$) and ignoring the wave packet spreading
with $t$. 
\par 
For finite but small $a$ one similarly finds
\begin{equation}
    (\Delta y_2)^2|_{a}=(\Delta y_2)^2|_{a\to 0}
    (1+2 a^2 \delta^\prime)\ , 
\end{equation}
where
\begin{equation}
\delta^\prime=\frac{\sigma_+^2+\sigma_-^2}{12}
\left[2\frac{1+\sigma_+^2\sigma_-^2{t^2\over m^2}}
{\left(1+\sigma_+^4{t^2\over m^2}\right)\left(1+\sigma_-^4{t^2\over m^2}\right)}
-\frac{1+\left(\frac{2\sigma_+\sigma_-}{\sigma_+^2+\sigma_-^2}\right)^2}
{1+\sigma_+^2\sigma_-^2{t^2\over m^2}}\right]\ ,
\end{equation}
is never negative.
\par 
According to the results of the two last paragraphs, the spreading of the vertical
right-side momentum, $\Delta k_2$, in coincidence events with a left-side particle
passing through the $s_L=2a$ wide slit, is indeed affected by the physical presence
of an equal $2a$-wide slit on the right side. 
Its presence filters a narrower (in $y_2$) right-moving wave packet and this 
translates into a $\Delta k_2$ which is larger than in the case of removing
(or making $s_R\gg 2a$ for) the right-side slit. 
This conclusion is in agreement with the analysis made by Short
\cite{Short} of the optical experiment performed by Kim and Shih \cite{KS}.

\section{Conclusions}
\label{summary}
We have reconsidered Popper's test using the standard quantum mechanical formalism
and, consequently, using a wave packet for the source ---or, equivalently, for the
initial two-particle system--- which satisfies Heisenberg's principle, $\Delta (k_1 + k_2)
\Delta \left[ (y_1 + y_2)/2 \right] \ge 1/2$. This contrasts with the original
Popper's proposal involving a fixed source and therefore subjected to the criticisms
raised by several authors \cite{BS,CL,Peres,Short,Qureshi1,Qureshi2}. 
In spite of this and contrary to the claims of some of these authors 
\cite{BS}--\cite{Qureshi2}, we find that Popper's test can be
conclusive in that a narrowing of the left-side slit increases $\Delta k^2_2$ of the
freely right-moving particle in coincidence detections. In other words, the
qualitative behavior of $\Delta k^2_2$ that Popper attributes to standard quantum
mechanics remains valid with our improved treatment of the initial state. In agreement
with a related analysis by Short \cite{Short}, we find however that the other aspect
of Popper's proposal gets modified by our analysis; namely, $\Delta k^2_2$ necessarily
increases when a second slit is {\it really} and symmetrically inserted on the
right-side of the setup. 
\par
Our analysis shows that, to some extend, Popper's test can indeed be conclusive to
discriminate between his own approach and the standard version of quantum mechanics. 
The latter turns out to be favored by recent optical experiments, which are somehow
related to the original proposal \cite{KS,Sergienko,Zeilinger,Dopfer}. Quantum
non-locality, a key question in all these discussions, is nowadays firmly established. 
Recent experiments tend to falsify Popper's approach but his 
understanding of quantum mechanics as early as in 1934 
\cite{Popper34} is quite remarkable.

\textit{Note added:}
After publication of the present paper in the Proceedings of the
Fundamental Physics Meeting ``Alberto Galindo''
we have received an improved version of Qureshi's paper in 
Ref.~\cite{Qureshi2}.
The state discussed in this new version coincides with our 
Eqs.~(\ref{xx1}) and (\ref{kk}) once we define
$\sigma_+^2=1/4\Omega_0^2$ and $\sigma_-^2=4\sigma^2$.

\paragraph{Acknowledgements}
This work is  partly supported by the Ramon y Cajal program (R.E.),
the Ministerio de Ciencia y Tecnolog\'{\i}a and FEDER, BFM-2002-02588,
and the EU, HPRN-CT-2002-00311, EURIDICE network.

\end{document}